\documentclass{article}
\usepackage{spconf,amsmath,epsfig}
\usepackage{pifont}

\usepackage{subfig}
\usepackage{booktabs}
\usepackage{array}
\usepackage{float}
\usepackage{amssymb}
\usepackage{color}
\usepackage{multirow,cite}
\usepackage[multiple]{footmisc}

\let\OLDthebibliography\thebibliography
\renewcommand\thebibliography[1]{
  \OLDthebibliography{#1}
  \setlength{\parskip}{0pt}
  \setlength{\itemsep}{0pt plus 0.3ex}
}

\begin{document}\sloppy

\def\x{{\mathbf x}}
\def\L{{\cal L}}

\title{Description on IEEE ICME 2024 Grand Challenge: Semi-supervised Acoustic Scene Classification under Domain Shift}

\name{
\parbox{\linewidth}{\centering
Jisheng Bai{\rm\textsuperscript{1,2,3}},
Mou Wang{\rm\textsuperscript{4}},
Haohe Liu{\rm\textsuperscript{5}}, 
Han Yin{\rm\textsuperscript{1}},
Yafei Jia{\rm\textsuperscript{1}}, 
Siwei Huang{\rm\textsuperscript{1}}, \\
Yutong Du{\rm\textsuperscript{1}},
Dongzhe Zhang{\rm\textsuperscript{1}},
Dongyuan Shi{\rm\textsuperscript{3}},
Woon-Seng Gan{\rm\textsuperscript{3}},\\
Mark D. Plumbley{\rm\textsuperscript{5}}, 
Susanto Rahardja{\rm\textsuperscript{1}},
Bin Xiang{\rm\textsuperscript{2}},
Jianfeng Chen{\rm\textsuperscript{1}}
}
}
\address{
	\textsuperscript{1} School of Marine Science and Technology, Northwestern Polytechnical University, Xi'an, China\\
     \textsuperscript{2} Xi'an Lianfeng Acoustic Technologies Co., Ltd., China\\
    \textsuperscript{3} School of Electrical \& Electronic Engineering, Nanyang Technological University, Singapore\\
    \textsuperscript{4} Institute of Acoustics, Chinese Academy of Sciences, Beijing, China\\
     \textsuperscript{5} University of Surrey, UK
}

\maketitle

\begin{abstract}
Acoustic scene classification (ASC) is a crucial research problem in computational
auditory scene analysis, and it aims to recognize the unique acoustic characteristics of an environment. 
One of the challenges of the ASC task is the domain shift between training and testing data.
Since 2018, ASC challenges have focused on the generalization of ASC models across different recording devices. 
Although this task, in recent years, has achieved substantial progress in device generalization, the challenge of domain shift between different geographical regions, involving discrepancies such as time, space, culture, and language, remains insufficiently explored at present.
In addition, considering the abundance of unlabeled acoustic scene data in the real world, it is important to study the possible ways to utilize these unlabelled data. 
Therefore, we introduce the task \textit{Semi-supervised Acoustic Scene Classification under Domain Shift} in the ICME\footnote{https://2024.ieeeicme.org/} 2024 Grand Challenge\footnote{https://ascchallenge.xshengyun.com/}.
We encourage participants to innovate with semi-supervised learning techniques, aiming to develop more robust ASC models under domain shift. 
\end{abstract}
\section{Introduction}
\label{sec:intro}

Acoustic scene classification (ASC) aims to identify an acoustic scene among the predefined classes in the environment, using signal processing and machine learning methods, such as squares, streets, and restaurants.
ASC systems can potentially benefit many applications, such as wearable devices, robotics, and smart home devices \cite{barchiesi2015acoustic, ding2023acoustic, jeong2022cochlscene}.

The Detection and Classification of Acoustic Scenes and Events (DCASE) Challenges\cite{mesaros2017detection} have been held in recent years and have generated wide attention. As one of the core tasks of the DCASE Challenge, ASC has attracted a lot of interest and has been widely investigated. 
In particular, deep learning algorithms have emerged as the predominant approach, significantly enhancing ASC performance \cite{heittola2020acoustic, mesaros2017detection, Schmid2023, wan2019ciaic}.
Deep learning-based ASC methods typically demand substantial data to achieve leading performance, giving rise to two critical considerations in the development of deep learning approaches for ASC: domain shift and scarcity of labeled data\cite{singh2021prototypical}.
\begin{figure}[t!]
	\centering
	\includegraphics[width=8cm]{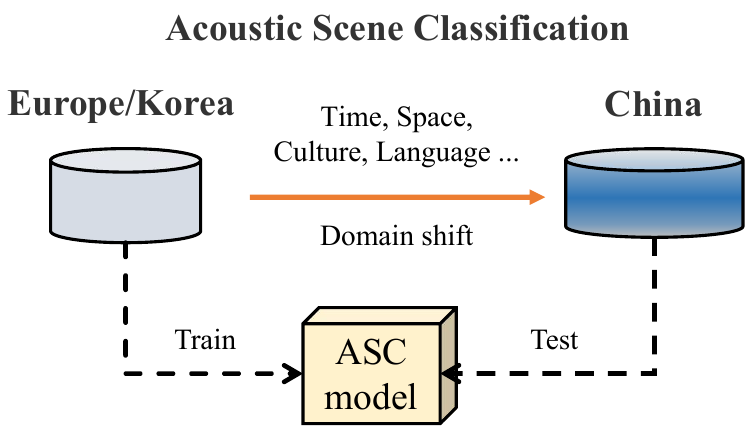}
	\caption{The domain shift problem in acoustic scene classification.}
	\label{fig:ASC domain}
\end{figure}
The domain shift problem is general in deep learning-based ASC systems since there is often a mismatch between training and testing data \cite{heittola2020acoustic}.
Previous works have studied the mismatch of devices and cities in ASC \cite{mesaros2018multi, tan2024acoustic}. 
The TAU Urban Acoustic Scenes (UAS) datasets, collected from $12$ European cities across $11$ devices, have been used for studying the mismatch problems in ASC\cite{mesaros2019acoustic}. 
After that, DCASE 2020 Task 1A\footnote{https://dcase.community/challenge2020/task-acoustic-scene-classification\#subtask-a}, titled \textit{Acoustic Scene Classification with Multiple Devices}, was introduced to study the device mismatch issues. 
However, the concentration of recording cities in Europe results in a dataset where people share more similar living environments compared to other continents like Asia and Africa.
In our initial experiments, we trained an ASC model using the TAU UAS 2020 Mobile development dataset and evaluated its performance on 14,000 recordings collected in China. 
The pre-trained model achieved a mean accuracy score of $14\%$, indicating a significant decline in ASC performance.
This highlights the challenge of applying existing datasets to environments with broader differences in culture, language, and infrastructure.

Deep learning methods have been successfully applied in ASC and achieved excellent results\cite{ding2023acoustic, kong2020panns, chen2022hts, gong2021ast}.
However, deep learning methods require a large amount of data for training, and building a large-scale acoustic scene dataset is time-consuming and labor-intensive. 
Semi-supervised learning, as an effective machine learning approach, can leverage both labeled and unlabeled data, reducing the dependence on labeled data \cite{serizel2018large}.
To the best of our knowledge, we are the first to propose an international challenge specifically exploring semi-supervised learning in ASC.

To delve deeper into the domain shift and semi-supervised research topics, we proposed \textit{Semi-supervised Acoustic Scene Classification under Domain Shift} in the ICME 2024 Grand Challenge. 
We collected a large-scale dataset from $10$ common acoustic scenes across $22$ Chinese cities using $3$ industrial recording devices in 2023, and named it the \textit{Chinese Acoustic Scene}~(CAS) 2023 dataset.
To facilitate the challenge, we present a development dataset derived from the CAS 2023 dataset, comprising $4.8$ hours of labeled data and $19.3$ hours of unlabeled data. 
We encourage participants to utilize this dataset for the development of effective semi-supervised techniques. 
Regarding the evaluation dataset, we will provide approximately $3$ hours of data for evaluations at the final stage of the challenge.
This dataset incorporates recordings from cities that have not been previously seen, allowing for an exploration of the domain shift issue.

The organization of this paper is arranged as follows.
Section 2 introduces the dataset, section $3$ introduces the baseline system, and section 4 concludes this paper.

\section{Dataset}
\label{sec:dataset}
In this section, we will introduce the CAS dataset, development, and evaluation dataset of the ICME ASC challenge.

\subsection{The Chinese Acoustic Scene Dataset}
The CAS 2023 dataset is a large-scale dataset that serves as a foundation for research related to environmental acoustic scenes. 
It includes $10$ common acoustic scenes, with a total duration of over $130$ hours. 
Each audio clip is $10$ in length with a 48 kHz sampling rate and accompanying metadata of location and timestamp information. 
The dataset was collected by members of the \textit{Joint Laboratory of Environmental Sound Sensing at the School of Marine Science and Technology, Northwestern Polytechnical University}. 
The data collection period spanned from April 2023 to September 2023, covering $22$ different cities across China.

\subsubsection{Recording Device}
\begin{figure}[t!]
	\centering
	\includegraphics[width=9cm]{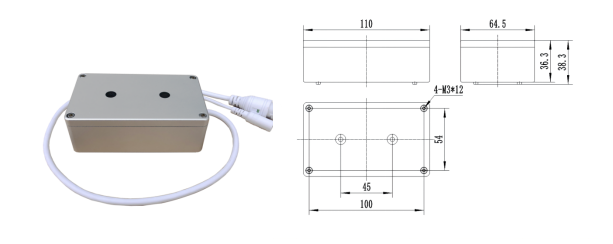}
	\caption{The recording device of the CAS dataset. The left side of the figure shows the physical representation of the device, and the right side displays the relevant dimensional parameters of the device, measured in millimeters.}
	\label{fig:2BE1}
\end{figure}

The CAS 2023 dataset was collected using XS-SN-2BE1 manufactured by \textit{Xi'an Lianfeng Acoustic Technologies Co., Ltd}. 
Fig. \ref{fig:2BE1} shows the physical and dimensional representation of XS-SN-2BE1.
It is a binaural sound monitoring device with the TCP/IP communication protocol for data acquisition.
It can operate in temperatures ranging from $-25^{\circ}C$ to $55^{\circ}C$, and it meets the IP67 standards for waterproof and dustproof.
It supports a sampling rate of up to 125kHz with low power consumption requirements. 
This device can be further developed with SDKs or other software platforms to enable sound detection or classification in specific scenarios.
More details of the recording device can be found on the official website\footnote{https://www.lfxstek.com}.

\subsubsection{Recording Procedure}
To ensure quality and diversity while recording the acoustic scene data, we follow the specific guidelines: \\
1. Original recordings have two channels, each with a duration of 10 minutes.\\
2. The recording person can carry the device during the recording process to simulate real-world scenarios but should avoid large movements like running or jumping.\\
3. The recording person should avoid any impact on the device during recording. \\
4. The recording device automatically records the timestamp. The recording person is also required to note the time for cross-verification, and also note down the the city of recording to provide additional location information.

\subsubsection{Recording Annotation}
To facilitate the annotation procedure, we developed an internal web platform that supports tasks such as uploading recordings and metadata, annotating, inspecting, and storing data.
The recording and annotation tasks were undertaken by researchers with at least one year of acoustic research experience. 
Throughout the annotation process, recordings containing personal identifiable information were carefully reviewed and deleted.

\subsection{Challenge Datasets}
The ICME ASC challenge dataset consists of development and evaluation datasets, all derived from the CAS 2023 dataset\footnote{https://zenodo.org/records/10616533}.
The development dataset is about 24 hours including the recordings from 8 cities.
We randomly provided scene labels for 20\% of the data in the development dataset to allow participants to develop effective semi-supervised methods.
The $10$ acoustic scene classes and the number of labeled recordings for each scene are shown in Table \ref{table:dev_label}.
In the evaluation dataset, data are selected from 12 cities, with 5 unseen cities specifically chosen to provide a more comprehensive evaluation of submissions under domain shift.

\begin{table}[h]
\centering
\caption{The number of labeled recordings for each scene in the development dataset.}
\label{table:dev_label}
\begin{tabular}{ccc}
\hline
\textbf{Scene}    & \textbf{Num of labeled recordings} \\ \hline
Bus               & $188$                                \\
Airport           & $220$                               \\
Metro             & $209$                               \\ 
Restaurant        & $173$                                \\
Shopping mall     & $147$                              \\  
Public square     & $174$                               \\ 
Urban park        & $148$                              \\  
Traffic street    & $143$                              \\  
Construction site & $173$                               \\
Bar               & $165$                                \\ \hline
Total             & $1740$                               \\ \hline
\end{tabular}
\end{table}

\section{Baseline}
\label{sec:baseline}
\begin{figure}[t!]
	\centering
	\includegraphics[width=8.5cm]{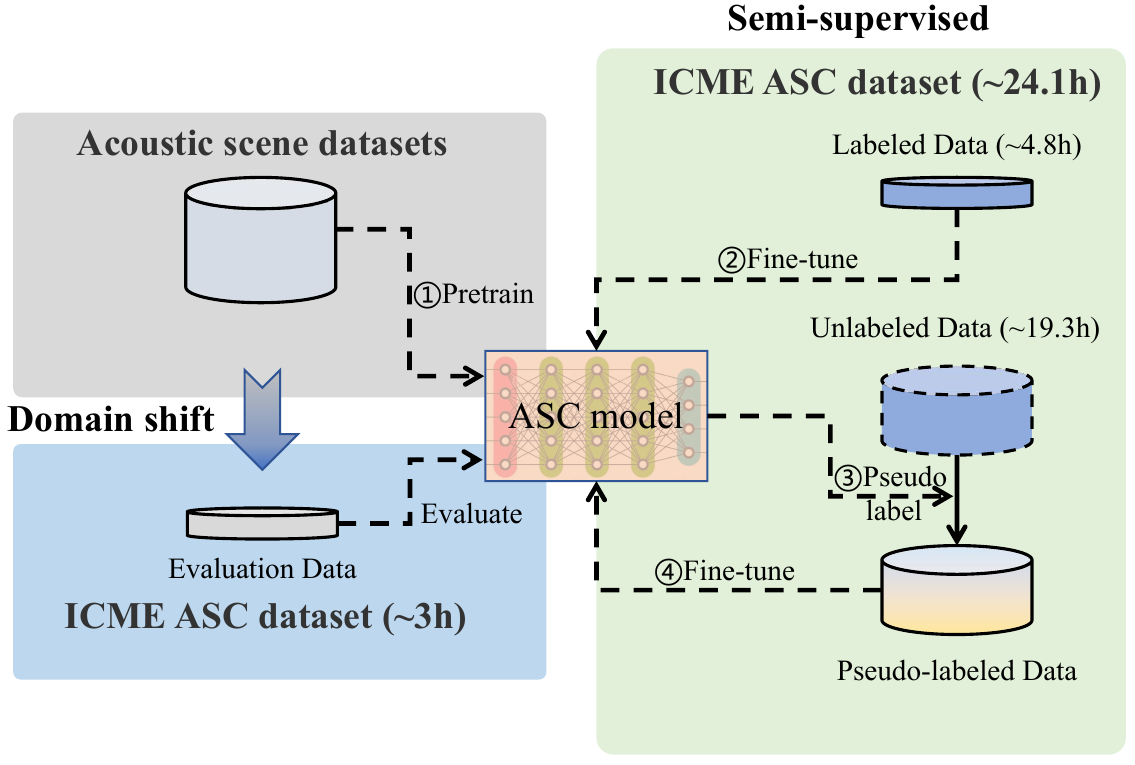}
	\caption{The pipeline of the challenge baseline.}
	\label{fig:pipeline}
\end{figure}
\subsection{Overview}
The baseline system for the ICME 2024~\textit{Semi-supervised Acoustic Scene Classification under Domain Shift} challenge is based on a semi-supervised framework with pre-trained attention-based model\footnote{https://github.com/JishengBai/ICME2024ASC}.
The pipeline of the baseline system is shown in Figure~\ref{fig:pipeline}.
First, we trained the ASC model using the TAU UAS 2020 Mobile development dataset, utilizing it as the pre-trained ASC model. 
Second, we performed fine-tuning on the pre-trained ASC model using the labeled data from the development dataset. 
Third, we used the ASC model to assign pseudo labels to the unlabeled data within the development dataset. 
Fourth, we utilized the pseudo-labeled data for further fine-tuning on the SE-Trans, resulting in the final ASC model used for evaluation.

\subsection{Baseline Model Architecture}
The architecture of the ASC model used in the baseline is the Squeeze-and-Excitation and Transformer (SE-Trans), which achieves state-of-the-art ASC performance in cross-task models of
environmental sound recognition \cite{bai2022squeeze}. 
In the baseline of the ICME ASC challenge, we adopt the best configuration of the SE-Trans from \cite{bai2022squeeze}.
The model consists of two SE blocks and one Transformer encoder, and each SE block consists of two convolutional layers with the same channels and kernel sizes of $3\times 3$. 
The number of channels of the first and second SE blocks is $64$ and $128$, respectively. 
An average pooling layer is applied after each SE block with kernel sizes of $2\times 2$. 
For the Transformer encoder, we set $8$ as the number of heads, $1$ as the number of layers, and $32$ as the number of units of fully connected layers. 
The final output is obtained through max aggregation across time frames and a fully connected layer.

\subsection{Experimental Setups}
We used log mel spectrograms as the input features. 
First, all recordings were resampled to 44,100 Hz. 
The short-time Fourier transform (STFT) with a Hanning window of $40$ ms and a hop size of $20$ ms was used to extract the spectrogram. 
We applied $64$ mel-filter bands on the spectrograms followed by a logarithmic operation to calculate log mel spectrograms.
Each log mel spectrogram has a shape of $500\times64$, where $500$ is the number of time frames and $64$ is the number of frequency bins.
During the fine-tuning of the baseline, we used Adam optimizer with a learning rate of $0.001$ and a batch size of $32$.

\subsection{Evaluation}
The evaluation metric for this competition is macro-average accuracy, which was commonly used in previous ASC challenges \cite{mesaros2017detection, heittola2020acoustic}.
It is calculated as the average of class-wise accuracies:
\begin{equation}
\text{Accuracy} = \frac{1}{N} \sum_{i=1}^{N} \text{Accuracy}_{i},
\end{equation}
where $N$ is the number of classes, and $\text{Accuracy}_{i}$ is the accuracy for class $i$.

\subsection{Baseline Results}
The ASC performance of the baseline system for each scene on the evaluation dataset is shown in Table \ref{table:baseline_performance}.
The model was trained and tested using the same random seed to ensure reproducibility of the results.
The baseline system achieves a macro-average accuracy score of $59\%$ on the evaluation dataset.
We note that the baseline achieves higher accuracies for \textit{Metro} and \textit{Bar} compared to other scene classes. 
The most challenging scene class to recognize is \textit{Public square}, with the lowest accuracy of $29\%$.
We assume that the higher accuracies for \textit{Metro} and \textit{Bar} may be attributed to the presence of unique sound events, such as running sounds of metro vehicles and music. Conversely, the lower accuracy for \textit{Public square} indicates that it may have similar acoustic content with other scene classes.

\begin{table}[h]
\centering
\caption{ASC performance of the baseline system for each scene on the evaluation dataset.}
\label{table:baseline_performance}
\begin{tabular}{ccc}
\hline
\textbf{Scene}    & \textbf{Accuracy} \\ \hline
Bus               & $40$\%                                \\
Airport           & $55$\%                              \\
Metro             & $90$\%                               \\ 
Restaurant        & $69$\%                                \\
Shopping mall     & $51$\%                              \\  
Public square     & $29$\%                               \\ 
Urban park        & $46$\%                              \\  
Traffic street    & $65$\%                              \\  
Construction site & $68$\%                               \\
Bar               & $87$\%                                \\ \hline
Average             & $59$\%                               \\ \hline
\end{tabular}
\end{table}

\section{Conclusion}
\label{sec:conclusion}

This paper introduces the IEEE ICME 2024 Grand Challenge: Semi-supervised Acoustic Scene Classification under Domain Shift. 
The challenge aims to explore semi-supervised learning methods that can be integrated into ASC systems, addressing the challenges of data collection time constraints and domain generalization in ASC models. 
To organize the challenge, we collected the CAS 2023 dataset with $10$ acoustic scenes across $22$ Chinese cities, providing an additional dataset with different geographical and cultural backgrounds for the research community.
We also release the baseline with a semi-supervised framework to encourage the participants to develop innovative semi-supervised methods.
We will provide an evaluation dataset that includes $5$ unseen cities compared to the development dataset to evaluate the submissions under domain shift conditions.

\section{Acknowledgement}
The recording devices for the dataset and the prizes for ICME ASC GC winners are sponsored by Xi'an Lianfeng Acoustic Technologies Co., Ltd.
We acknowledge the support provided by the China Scholarship Council (CSC) during a visit of Jisheng Bai to Nanyang Technological University.
This research was partly supported by Engineering and Physical Sciences Research Council (EPSRC) Grant EP/T019751/1 “AI for Sound”, and a PhD scholarship from the Centre for Vision, Speech and Signal Processing (CVSSP) at the University of Surrey and BBC R\&D.

\bibliographystyle{IEEEbib}
\bibliography{icme2023template}

\end{document}